# Uses and Gratifications of Alternative Social Media: Why do people use Mastodon?

Kijung Lee[1] and Mian Wang[1]

[1] University of Cincinnati, Cincinnati OH 45221, USA
kijung.lee@uc.edu

**Abstract.** The primary purpose of this investigation is to answer the research questions; 1) What are users' motivations for joining Mastodon?; 2) What are users' gratifications for using Mastodon?; and 3) What are the primary reasons that the users continue to use Mastodon? We analyzed the collected data from the perspective of the Uses and Gratifications Theory. A questionnaire was designed to measure the opinions of Mastodon users from 15 different Mastodon instances. We examined 47 items through exploratory factor analysis using principal components extraction with Varimax with Kaiser Normalization. The results extracted 7 factors of gratification sought (expectation) and 7 factors of gratification obtained. We discovered that the primary reason that the users join and use Mastodon is the ease of controlling and sheltering users' information from data mining. The findings of the gratification sought structure are similar to findings of the gratification obtained structure, and the comparison between the two groups of data suggests that users are satisfied with the ongoing use of Mastodon.

## 1 Introduction

The emergence of social media radically changed how people communicate and interact with others. Using a collection of platforms, e.g., Facebook, Twitter, and Youtube, to name a few, social media users share ideas and collaborate with the community of others. However, the often-mentioned pure function of social media is sometimes overshadowed by dysfunctional issues, essentially summarized as information control. The first and foremost issue of information control is privacy. Most social media users are concerned that the information they share on social media platforms is not protected for privacy, especially how their private information is collected and used for targeted advertising (Haddadi, Hui, Henderson, & Brown, 2011; Hayes, Brinson, Bott, & Moeller, 2021; Zhang, Guerrero, Wheatley, & Lee, 2010). On the other hand, the failure of the service providers to moderate appropriate content across social platforms contributed to the spread of misinformation. While content moderation may be accepted to filter out offensive messages, it is unpleasant for some people due to its potential bias and lack of efficacy (Gillespie, 2020; Myers West, 2018). In addition to privacy and misinformation issues, mainstream social media have been criticized for censorship, political neutrality, user control, and malicious activities (Casilli & Tubaro, 2012).

Many of the mentioned issues of mainstream social media contributed to users' transition to alternative social media (ASM henceforth). ASM emerged in 2010th, led by a small number of startups, including Diaspora. One of the primary differences between mainstream social media services (e.g., Facebook and Twitter) and ASM is the user's control over their data. In other words, the decentralized nature of ASM enables the users to manage their user-created-contents thanks to the lack of a single service provider. An example of users' transition from mainstream social media to ASM is the recent cyber-migration from Twitter to Mastodon in October 2019. The Twitter account of Sanjay Hegde, a Supreme Court lawyer, was suspended because of the excessive content he posted, which triggered this movement. Twitter users were unhappy with the censorship policies and content verification (Ranipeta, 2019). Mastodon garnered over 1.1 million during that time (Parker, 2012).

As the largest and most popular ASM, Mastodon is one representative and notable example. Drawing on data from the Federation hub (Parker, 2012), Mastodon is the largest and most popular platform. It was created in 2016 by a German programmer, Eugen Rochko, to provide users with their "own space on the Internet." As a microblogging service, Mastodon is a decentralized web platform (Raman, Joglekar, Cristofaro, Sastry, & Tyson, 2019), installed on GNU Social (Kenlon, 2017), to host their servers and create their private networks. In line with Mastodon's slogan, "Giving social networking back to you," Mastodon decentralized its network using many independent local servers. More specifically, Mastodon lets users set up their local server (an instance) with complete administration and allows users to register and interact with each other. The primary benefit is security and authentication compared to on-premises and cloud applications (Parker, 2012). Many people see and use Mastodon as an alternative to Twitter because they both have similar user interfaces and built-in mechanisms.

With less restrictive content moderation by the service providers, the ASM may potentially provide a forum with "free" speech and improved data privacy. Freedom of expression is not a simple idea to implement as a policy of service because public speaking, in the most part, is governed by the law rather than the company policy or the lack of it. However, as the ASM are not managed by a single entity, they have their version of "content moderation," although it is more like a formal agreement among the members of a particular group. In practice, there are ways that users can post content without censorship. This is one of the major reasons that users prefer ASM over traditional social media. However, the unfamiliarity with the decentralized structure of how ASM work, in addition to the less appealing user interface, contributes to the low number of active users.

Theoretical examination of ASM… uses and gratifications

The primary purpose of this investigation is to answer the research questions; *1) What are users' motivations for joining Mastodon?; 2) What are users' gratifications for using Mastodon?; and 3) What are the primary reasons that the users continue to use Mastodon?* We analyzed the collected data from the perspective of the Uses and Gratifications (U&G henceforth) Theory. The rest of the paper is organized into the following sections; The Literature Review section describes the Uses and Gratifications Theory and how the theory can be applied in the context of the analyses of Mastodon users; In the Methods section, we describe procedures and participants, instrument design, and statistical approach to answer the research questions; In the Results and Analysis section, we provide the results of the analyses and our interpretations of how the results answer the research questions; The we discuss the implications, contributions, and limitations and concludes the paper in the Discussion and Conclusion section.

## 2 Literature Review

Perspectives to explicate the transition from one form of media to a different form of media are often described in terms of how one offers the means for communication to marginalized people (Atton, 2008). In other words, alternative media are not only defined by the content that they deliver but also by how people can participate in the production of messages. In the traditional media, the distinction seems clear from newspaper to radio, from radio to television, and from television to web 2.0 media, that people’s choice of one media over another provided them with more power and opportunity to take part in the message production. Uses and Gratifications Theory is a theoretical framework to show why and how people select and use specific media to satisfy their needs.

### 2.1 Uses and Gratifications Theory

In the early 1960s, the U&G Theory was used to measure people's short-term effects on mass media campaigns (Blumler, 1979). Early theorists have applied the U&G perspective in the context of various mass media such as television and electronic bulletins (Eighmey & McCord, 1998). For example, Schramm’s  (Schramm, 1965) research about children seeking out television and collecting data on how TV affects them is a classical case. However, the work of Blumler and Katz and Blumler  (Katz & Blumler, 1974) changed researchers' perspectives to focus on what people do instead of the media's influence or impact on the individual. That resulted in the audiences becoming active objects where they could select any media as passive objects to satisfy their needs. This basic assumption also helps people understand the reasons for selecting different types of media. U&G Theory successfully distinguished itself from early mass communication theories and became a new form of the functionalist perspective on mass media communication (Luo, 2002). Katz and Blumler (Katz & Blumler, 1974) have found thirty-five social and psychological needs and divided them into five different basic categories: "Cognitive", "Affective", "Integrative", "Social Integrative", and "Escape". Furthermore, more and more gratifications have been discovered and used in studies such as: "Information Seeking", "Pass Time", "Entertainment", "Relaxation", "Communicatory Utility", "Convenience Utility", "Expression of Opinion", "Information Sharing", And "Surveillance/Knowledge About Others" (Whiting & Williams, 2013).

With the advancement of media technology, researchers have also paid attention to Web applications (e.g., SNSs). The interactive nature of this new type of media requires active user participation. Many researchers (LaRose, Mastro, & Eastin, 2001; Lee, 2004; Papacharissi & Rubin, 2000) have put the effort in the field to examine the expectations and gratifications that individuals seek and obtain. Understanding what motivates people to select social media has become the priority mission of U&G Theory. The results indicated that the users tend to use SNS to gain multiple gratifications. Users tend to use different platforms for receiving different gratifications. Quan-Haase et al. (Quan-Haase, Wellman, Witte, & Hampton, 2002) have addressed: They do not apply one single form of social media but a range of communication tools. All social media can be adapted to be a part of a user's communication repertoire. For example, people tend to use Facebook to communicate and socialize for real-life communication networks (Young, K., 2011), but people who use Twitter aim to ask questions, get support, or share advice (Grosseck & Holotescu, 2008). People are using different media types for different purposes and desires, and the U&G Theory can help researchers discover users' motivations and associated behaviors.

U&G theory examines why individuals use media to seek and obtain needs from specific media types as the most successful theoretical framework. In the review of U&G literature, many studies (Karimi, Khodabandelou, Ehsani, & Ahmad, 2014; Quan-Haase & Young, 2010; Quinn, 2016) only focus on mainstream social media, but few studies examined ASM from a U&G perspective. To address this deficiency in literature, the following research questions will be addressed from two aspects: Expectations for joining and gratifications for using Mastodon.

## 2.2 Motivations for using Mastodon

To date, Mastodon has 2.9 million users and 2,720 nodes (from the-federation.info), and this number is still going up during the COVID-19 pandemic. It has become the largest and the most popular ASM on the Internet and has also raised users' expectations of joining accordingly. Quan-Haase et al. (Quan-Haase & Young, 2010) have addressed that gratifications sought (also often referred to as "needs" or "motives") refer to those gratifications that audience members expect to obtain from a medium before they have come into contact with it. The users want to join Mastodon because they seek specific needs and use Mastodon to fulfill their expectations. This helps us find what factors influence Mastodon adoption and why users join. It helps us to find what factors influence adoption and why most users choose to join Mastodon. No studies have systematically investigated gratifications sought (expectations) from the Mastodon. Understanding the gap between these two types of gratifications is vital for analyzing how different audience members use various media kinds (Palmgreen & Rayburn, 1979). Hence, to explore the expectation for joining Mastodon, the first research question is the following:

Research Question 1: What are users' motivations for joining Mastodon?

### 2.3 Gratifications of using Mastodon

As many researchers (LaRose et al., 2001; Levy & Windahl, 1984; Palmgreen, Wenner, & Rayburn, 1980) have addressed, the gratifications obtained insight into what motivates the continued use of the medium. Ongoing use of one type of media implies that gratifications sought are reinforced by gratifications obtained. Although many studies examined gratifications obtained, e.g., (DiMicco et al., 2008; Joinson, 2008; Quan-Haase & Young, 2010), for mainstream media, no studies have systematically examined gratifications obtained from Mastodon. The gratifications obtained from using Mastodon will present reasons that support users for ongoing use of the platform. To better understand user's satisfaction, the comparisons between gratifications sought and gratifications obtained are required. Hence, the following research question is formed:

Research Question 2: What are users' gratifications for using Mastodon?

Research Question 3: What are the primary factors that the users continue to use Mastodon?

## 3 Methods

This paper presents a small-scale exploratory investigation of Mastodon users' motivations for joining and using Mastodon. The Questionnaire Survey method was used. This study consists of an online questionnaire with nineteen questions. The online questionnaire applies the U&G perspective to fulfill the purpose of the investigation.

### 3.1 Procedure and Participants

All samples were obtained under research protocols reviewed and approved by the University of Cincinnati Institutional Review Board (IRB) committee. Participation was voluntary, and participants for the survey were recruited from fifteen different Mastodon instances. The study employed the use of SurveyMonkey to apply questionnaires to Mastodon users. The flexibility of question types is useful for schools and school social workers (Massat, McKay, & Moses, 2009). Respondents accessed the questionnaire through a SurveyMonkey link and first were asked to read the information sheet that informed them everything about this investigation. Then, they answered all subsequent questions based on their usage of Mastodon. Data collection took place between February 2021 and March 2021.

Two hundred fifty-four participants were initially recruited from fifteen different Instances of the Mastodon that have the most active user ratio (the highest ratio is 87%, and the lowest ratio is 9.2%). Due to missing data and an incomplete survey, the final sample size for answering research questions was one hundred fifty. Of this sample, 78% of the respondents were female, 17% were male, and 5% preferred not to tell. The age range of 18-24 had the most significant number of respondents (62.4%), followed by the age range of 25-34 (32.4%) and 35-44 (4.5%). The majority of participants had completed a bachelor's (43.5%) and a master's degree (27.8%). Most of the participants lived in China (61.33%) the and U.S. (12.44%).

### 3.2 Measurements

The online survey included three sections with nineteen questions: demographic information, gratification sought, and gratification obtained. In the demographic section (with four questions), respondents were asked to indicate their age, gender, education level, and location. The demographic information provides an overview of respondents' backgrounds. The gratification sought section (with 47 items), and the gratification obtained section (with the same 47 items) were two measures of Mastodon. Participants could rate each item measured with 5-point Likert scale, ranging from "STRONGLY DISAGREE" to "STRONGLY AGREE." The questionnaire contains thirty-two items that were created for indicating reasons people use mainstream social media (e.g., Twitter). Other fifteen items were extracted from the online conversion with Mastodon users, especially about privacy and free speech issues. The questionnaire is divided into three sections: "Basic Information", "Why do you join Mastodon?" and "Why do you use Mastodon?" Participants were asked to complete the phrases such as "I joined the Mastodon because I expected……" and "I use the Mastodon because……".

Table 1 below summarizes the commonly used phrases following "I joined the Mastodon because I expected……" and "I use the Mastodon because……" to measure the gratifications sought and gratifications obtained.

Table 1 Common phrases used to measure gratification sought and gratification obtained

| Factor | ID | Phrase |
|---|---|---|
| Social Escapism and Support | SE1 | Get away from what I am doing |
| | SE2 | Express negative emotions |
| | SE3 | Stirs me up |
| | SE4 | Make me feel less lonely. |
| | SE5 | Escape from reality. |
| | SE6 | Help me unwind. |
| | SE7 | Express my feelings without causing unnecessary social attention. |
| | SE8 | Show me how to get along with others. |
| | SE9 | Avoid someone online |
| | SE10 | Let out my emotions easily to others who will sympathize |
| | SE11 | Not to be alone. |
| | SE12 | Let others know I care about their feelings |
| | SE13 | I want to give myself enough entertainment. |
| Adult Content sharing | AS1 | Safe to share adult content; |
| | AS2 | share adult content photos; |
| | AS3 | share adult webpage links; |
| | AS4 | share adult content videos; |
| | AS5 | share my own adult contents; |
| | AS6 | share adult content audio clips; |
| | AS7 | Find dates |
| Adult Content Consuming | AC1 | Only for adult content |
| | AC2 | adult content that can only be posted on Mastodon for entertainment. |
| | AC3 | find adult only instances. |
| | AC4 | find links that contain adult content. |
| | AC5 | find adult videos. |
| | AC6 | find adult photos. |
| | AC7 | find adult audio clips. |
| Information Seeking | IS1 | I want to find someone similar to me. |
| | IS2 | I want to meet people with the same interests. |
| | IS3 | I believe it allows me to meet new people. |
| | IS4 | Learn about unknown things |
| | IS5 | Do research |
| | IS6 | Get new ideas |
| | IS7 | Learn about useful things |
| Socialization | S1 | my friend suggested it to me. |
| | S2 | Keep in touch with someone. |
| | S3 | I am looking for someone I know. |
| | S4 | all my friends are using Mastodon. |
| | S5 | Find a job |
| | S6 | Find a date |
| Privacy | P1 | decentralized platform cannot collect my personal information |
| | P2 | Not be found by people I do not know |
| | P3 | decentralized platform will prevent people who want stalk me |
| | P4 | Not be exposed on SNS |
| | P5 | None-decentralized SNSs cannot provide sufficient privacy settings; |
| Convenience | CO1 | Use it at anytime and anywhere |
| | CO2 | Use it easily |
| | CO3 | Use it conveniently |

# 4 Results and Analysis

## 4.1 Research Question 1: Motivations for using Mastodon

When examining what motivates Mastodon users to join Mastodon, 12 factors were identified based on factor analysis with eigenvalues greater than 1.0. However, after running a parallel analysis (Factor Analyzed = 47; n = 150; Calculated Eigenvalues (1.548) < Mean Eigenvalue (1.638954)), five factors were removed, and seven factors were retained (see Table 2).

Factor 1, "Social Escapism and Support", is comprised of thirteen items measuring how users expect Mastodon can provide a place to help them avoid real-world problems or get away from responsibilities (e.g., depression and difficulties). The initial eigenvalue is 4.7519, and this factor explains 22.448% of the total variance. The mean of these items was very high, suggesting that the items represented vital expectations. Two key expectations were " Express my feelings without causing unnecessary social attention" (Mean = 4.18) and "Let out my emotions easily to others who will sympathize" (Mean = 3.83). Factor 2 contains "Adult Content Sharing" items. It was the second-highest factor (eigenvalue 4.182, variance explained 9.595%) comprising of seven items measuring Mastodon Users' expectation to microblog services as an adult content sharing community. Two key expectations were "Safe to share adult content" (Mean = 3.1133) and "Want to share adult webpage links" (Mean = 2.3467). Factor 3, "Adult Content Consuming" (eigenvalue 4.0771, variance explained 6.776%), consists of seven items measuring the participants' expectation for entertainment on Mastodon. Two key expectations were "adult content that can only be posted on Mastodon for entertainment" (Mean = 2.28) and "Want to find links that contain adult content" (Mean = 2.14). "Information Seeking" was the fourth factor identified (eigenvalue 2.46026, variance explained 5.416%) and consists of six items measuring the participants' expectation for willingness to join a new platform and search for information of people and events. Factor 5, "Socialization" (eigenvalue 2.28028, variance explained 4.727%), includes four items measuring the participants' expectation for feeling involved with society and what is going on with others and friendship toward others. Factor 6, "Privacy" (eigenvalue 2.01714 variance explained 4.322%), comprises five items measuring the expectation of privacy protection online on Mastodon. Two key expectations were "None-decentralized SNSs cannot provide sufficient privacy settings" (Mean = 3.96) and "decentralized platform cannot collect my personal information" (Mean = 3.82). Factor 7, "Convenience" (eigenvalue 1.66315, variance explained 4.098%), consists of 3 items: "Use it conveniently"(Mean = 3.7733), "Use it easily"(Mean = 3.8667), and "Use it at anytime and anywhere"(Mean = 3.7533).

Table 2 Factor Loadings (Principal Components, Varimax with Kaiser Normalization, Suppress Small Coefficients: 0.3) of 47 Gratifications Sought (N 150)

| I joined the Mastodon because I expected: | mean | sd | factors | | | | | | |
|---|---|---|---|---|---|---|---|---|---|
| | | | 1 | 2 | 3 | 4 | 5 | 6 | 7 |

| | | | | | | | |
|---|---|---|---|---|---|---|---|
| Social Escapism and Support | | 3.36 | | | | | |
| | SE1 | 3.24 | 1.16 | 0.76 | | | |
| | SE2 | 3.71 | 1.14 | 0.64 | | | |
| | SE3 | 3.06 | 1.08 | 0.74 | | | |
| | SE4 | 3.63 | 1.08 | 0.70 | | | |
| | SE5 | 3.21 | 1.14 | 0.67 | | | |
| | SE6 | 3.10 | 0.95 | 0.66 | | | |
| | SE7 | 4.18 | 0.93 | 0.64 | | | |
| | SE8 | 2.63 | 1.02 | 0.62 | | | |
| | SE9 | 3.49 | 1.33 | 0.58 | | | |
| | SE10 | 3.83 | 0.99 | 0.48 | | | |
| | SE11 | 3.06 | 1.06 | 0.44 | | | |
| | SE12 | 3.18 | 1.03 | 0.42 | | | |
| | SE13 | 3.53 | 0.85 | 0.33 | | | |
| Adult Content sharing | | 2.32 | | | | | |
| | AS1 | 3.11 | 1.13 | | 0.52 | | |
| | AS2 | 2.14 | 0.99 | | 0.89 | | |
| | AS3 | 2.35 | 1.07 | | 0.77 | | |
| | AS4 | 2.13 | 0.99 | | 0.88 | | |
| | AS5 | 1.89 | 1.05 | | 0.87 | | |
| | AS6 | 2.09 | 0.98 | | 0.88 | | |
| | AS7 | 1.61 | 0.79 | | 0.48 | | |
| Adult Content Consuming | | 1.86 | | | | | |
| | AC1 | 1.53 | 0.75 | | | 0.54 | |
| | AC2 | 2.28 | 1.12 | | | 0.73 | |
| | AC3 | 1.53 | 0.75 | | | 0.60 | |
| | AC4 | 2.14 | 1.12 | | | 0.74 | |
| | AC5 | 1.78 | 0.85 | | | 0.88 | |
| | AC6 | 2.01 | 1.03 | | | 0.84 | |
| | AC7 | 1.78 | 0.80 | | | 0.87 | |
| Information Seeking | | 3.43 | | | | | |
| | IS1 | 3.42 | 0.97 | | | | 0.71 |
| | IS2 | 3.53 | 1.02 | | | | 0.70 |
| | IS3 | 3.62 | 0.95 | | | | 0.63 |
| | IS4 | 3.78 | 0.88 | | | | 0.56 |

| | | | | | | | | | | |
|---|---|---|---|---|---|---|---|---|---|---|
| | IS5 | 2.15 | 0.91 | | | | 0.54 | | | |
| | IS6 | 3.63 | 0.95 | | | | 0.52 | | | |
| | IS7 | 3.90 | 0.73 | | | | 0.44 | | | |
| Socialization | | 2.45 | | | | | | | | |
| | S1 | 2.67 | 1.38 | | | | | 0.78 | | |
| | S2 | 2.91 | 1.35 | | | | | 0.72 | | |
| | S3 | 2.50 | 1.36 | | | | | 0.78 | | |
| | S4 | 1.72 | 0.87 | | | | | 0.74 | | |
| Privacy | | 3.74 | | | | | | | | |
| | P1 | 3.82 | 0.88 | | | | | | 0.68 | |
| | P2 | 3.67 | 0.99 | | | | | | 0.68 | |
| | P3 | 3.64 | 0.91 | | | | | | 0.57 | |
| | P4 | 3.63 | 0.90 | | | | | | 0.57 | |
| | P5 | 3.96 | 0.79 | | | | | | 0.67 | |
| Convenience | | 3.80 | | | | | | | | |
| | CO1 | 3.75 | 1.01 | | | | | | | 0.69 |
| | CO2 | 3.87 | 0.84 | | | | | | | 0.80 |
| | CO3 | 3.77 | 0.91 | | | | | | | 0.74 |
| | Eigenvalue | | | 4.75 | 4.18 | 4.08 | 2.46 | 2.28 | 2.02 | 1.66 |
| | Variance explained (%) | | | 22.45 | 9.60 | 6.78 | 5.42 | 4.73 | 4.32 | 4.10 |

### 4.2 Research Question 2: Gratifications of using Mastodon

When examining what gratifications pushed the use Mastodon, 20 factors were identified based on the factor analysis with eigenvalues greater than 1.0. However, after running a parallel analysis (Factor Analyzed = 47; n = 150; Calculated Eigenvalues (1.548) < Mean Eigenvalue (1.638954)), 13 factors were eventually removed, and seven factors were eventually retained and explaining 57.382 of the variances. (see Table 3).

Factor 1, "Social Escapism and Support", comprises eleven Social escapism and support items. The factor measures how users obtained the gratifications from using Mastodon to avoid real-world problems or to get away from responsibilities. The initial eigenvalue is 4.8092, and the factor explains 27.258% of the total variance. Two key gratifications were "Escape from reality" (Mean = 3.3977) and " Express my feelings without causing unnecessary social attention" (Mean = 3.8295), which shows how users on Mastodon use the platform as a beneficial association. Factor 2, "Adult Content Consuming" (eigenvalue 4.2925, variance explained 10.463%), consists of seven items measuring user satisfaction with the particular entertainment on Mastodon. Factor 3, "Adult Content sharing" (eigenvalue 3.9416, variance explained 7.611%), consists of six items measuring participants' satisfaction with sharing adult content on Mastodon. Two key gratifications were "Able to share adult video and not be discovered" (Mean = 3) and "Safe to share adult content" (Mean = 3). "Socialization" was the fourth factor identified (eigenvalue 2.9030, variance explained 5.834%) and consists of five items measuring the extent to which participants felt involved with society and what is going on with others and friendship toward others. The critical gratification was "Able to keep in touch with someone" (Mean = 3.0114). Factor 5, "Information Seeking" (eigenvalue 2.3946, variance explained 4.851%) and consists of nine items measuring the participants' gratification for willingness to use a new platform, meet new people, and search for information. The Mean for these items was very high. Two key gratifications were "Learned about unknown things" (Mean = 3.8750) and "Let out my emotions easily to others who will sympathize" (Mean = 3.7159). Factor 6, "Privacy" (eigenvalue 2.0730, variance explained 4.435%) and consists five items measuring the gratifications of privacy online on Mastodon. The Mean for these items was also very high. The critical gratification was "None-decentralized SNSs cannot provide sufficient privacy settings" (Mean = 3.9886). As the last gratification factor, "Convenience" (eigenvalue 1.1624, variance explained 3.528%), consists of three items, with the item "Use it conveniently" (Mean = 4.0455), "Use it easily" (Mean = 3.8977), and "Use it at anytime and anywhere" (Mean = 3.7045) measuring the extent to which participants willing to join a new platform.

Table 3 Factor Loadings (Principal Components, Varimax Rotation) of 47 Gratifications Obtained (N 150)

| I use Mastodon because: | | mean | sd | Factors | | | | | | |
|---|---|---|---|---|---|---|---|---|---|---|
| | | | | 1 | 2 | 3 | 4 | 5 | 6 | 7 |
| Social Escapism and Support | | 3.48 | | | | | | | | |
| | SE5 | 3.40 | 1.09 | 0.80 | | | | | | |
| | SE1 | 3.51 | 1.05 | 0.79 | | | | | | |
| | SE4 | 3.52 | 0.97 | 0.78 | | | | | | |
| | SE3 | 3.43 | 1.06 | 0.74 | | | | | | |
| | SE11 | 3.45 | 0.97 | 0.69 | | | | | | |
| | SE6 | 3.39 | 0.96 | 0.68 | | | | | | |
| | SE2 | 3.83 | 1.06 | 0.66 | | | | | | |
| | SE8 | 3.06 | 1.04 | 0.62 | | | | | | |
| | SE7 | 3.83 | 1.06 | 0.54 | | | | | | |
| | SE9 | 3.48 | 1.13 | 0.47 | | | | | | |
| | SE13 | 3.40 | 1.02 | 0.38 | | | | | | |

| | | | | | | | | |
|---|---|---|---|---|---|---|---|---|
| Adult Content Consuming | | 2.10 | | | | | | |
| | AC6 | 2.34 | 1.18 | 0.87 | | | | |
| | AC4 | 2.20 | 1.07 | 0.85 | | | | |
| | AC1 | 2.47 | 1.18 | 0.84 | | | | |
| | AC5 | 2.05 | 1.02 | 0.82 | | | | |
| | AC7 | 2.09 | 1.01 | 0.82 | | | | |
| | AC2 | 1.98 | 1.06 | 0.74 | | | | |
| | AC1 | 1.55 | 0.76 | 0.47 | | | | |
| Adult Content sharing | | 2.60 | | | | | | |
| | AS4 | 3.00 | 1.14 | | 0.90 | | | |
| | AS6 | 2.33 | 1.00 | | 0.87 | | | |
| | AS2 | 2.41 | 1.07 | | 0.87 | | | |
| | AS1 | 2.38 | 1.06 | | 0.86 | | | |
| | AS3 | 2.49 | 1.05 | | 0.83 | | | |
| | AS5 | 3.00 | 1.14 | | 0.45 | | | |
| Socialization | | 2.19 | | | | | | |
| | S2 | 3.01 | 1.31 | | | 0.80 | | |
| | S1 | 2.48 | 1.36 | | | 0.79 | | |
| | S3 | 2.81 | 1.34 | | | 0.75 | | |
| | S4 | 1.76 | 0.98 | | | 0.75 | | |
| | S5 | 1.57 | 0.74 | | | 0.53 | | |
| | S6 | 1.53 | 0.71 | | | 0.48 | | |
| Information Seek-ing | | 3.46 | | | | | | |
| | IS3 | 3.70 | 0.86 | | | | 0.68 | |
| | IS6 | 3.64 | 0.90 | | | | 0.55 | |
| | IS4 | 3.88 | 0.77 | | | | 0.47 | |
| | IS2 | 3.69 | 0.94 | | | | 0.46 | |
| | IS5 | 2.03 | 0.99 | | | | 0.45 | |
| | IS2 | 3.34 | 1.00 | | | | 0.43 | |
| | IS7 | 3.80 | 0.79 | | | | 0.40 | |
| Privacy | | 3.79 | | | | | | |
| | P1 | 3.67 | 0.87 | | | | | 0.71 |
| | P5 | 3.99 | 0.73 | | | | | 0.65 |
| | P3 | 3.76 | 0.83 | | | | | 0.65 |
| | P2 | 3.74 | 0.94 | | | | | 0.65 |

| | | | | | | | | | | |
|---|---|---|---|---|---|---|---|---|---|---|
| | P4 | 3.80 | 0.86 | | | | | | 0.54 | |
| Convenience | | 3.88 | | | | | | | | |
| | CO3 | 4.05 | 0.87 | | | | | | | 0.84 |
| | CO2 | 3.90 | 0.87 | | | | | | | 0.79 |
| | CO1 | 3.70 | 1.12 | | | | | | | 0.74 |
| Eigenvalue | | | | 4.81 | 4.29 | 3.94 | 2.90 | 2.39 | 2.07 | 1.16 |
| Variance explained (%) | | | | 27.26 | 10.46 | 7.61 | 5.83 | 4.85 | 4.44 | 3.53 |

### 4.3 Research Question 3: Gratification Sought and Gratifications Obtained

In Tables 2 and 3, both significant factors related to gratifications sought (G.S.) and gratifications obtained (G.O.) have been summarized individually. To better understand the G.S. and G.O.'s differences, all the factors have been sorted by mean (see Table 4), and a series of paired t-tests (see Table 5) have been applied. Comparing the mean differences between G.O. and G.S. were run to assess differences individually. From the seven pairs measured, three pairs (e.g., Adult Content Sharing; Adult Content Consuming; Social Escapism and Support) had statistically significant mean differences between G.O. and G.S., and four pairs (e.g., Socialization; Privacy; Convenience; Information Seeking) had no statistically significant mean differences. Among all the pairs, six pairs (e.g., Adult Content Sharing; Adult Content Consuming; Social Escapism and Support; Socialization; Privacy; Convenience; Information Seeking) have a positive variance (expectations exceeded).

Table 4 Factors sorted by Mean

| Factors | Gratification Sought | Mean | Gratifications Obtained | Mean |
|---|---|---|---|---|
| 1 | Convenience | 3.80 | Convenience | 3.88 |
| 2 | Privacy | 3.74 | Privacy | 3.79 |
| 3 | Information Seeking | 3.43 | Social Escapism and Support | 3.48 |
| 4 | Social Escapism and Support | 3.36 | Information Seeking | 3.46 |
| 5 | Socialization | 2.45 | Adult Content sharing | 2.60 |
| 6 | Adult Content sharing | 2.32 | Socialization | 2.51 |
| 7 | Adult Content Consuming | 1.86 | Adult Content Consuming | 2.10 |

Table 5 Paired t Tests for Gratifications Sought and Gratifications Obtained

| Pair | Factor | Mean | Mean diff. | sd | t |
|---|---|---|---|---|---|
| 1 | Adult Content sharing | | | | |
| | GO | 2.60 | 0.32 | 0.37 | 2.07 |
| | GS | 2.28 | | | |
| 2 | Adult Content Consuming | | | | |
| | GO | 2.10 | 0.23 | 0.47 | 1.31 |
| | GS | 1.86 | | | |
| 3 | Social Escapism and Support | | | | |
| | GO | 3.48 | 0.10 | 0.32 | 1.09 |
| | GS | 3.38 | | | |
| 4 | Socialization | | | | |
| | GO | 2.51 | 0.07 | 0.36 | 0.37 |
| | GS | 2.45 | | | |
| 5 | Privacy | | | | |
| | GO | 3.79 | 0.05 | 0.20 | 0.51 |
| | GS | 3.74 | | | |
| 6 | Convenience | | | | |
| | GO | 3.88 | 0.05 | 0.15 | 1.11 |
| | GS | 3.80 | | | |
| 7 | Information Seeking | | | | |
| | GO | 3.46 | 0.03 | 1.06 | 0.00 |
| | GS | 3.43 | | | |

# 5 Discussion and Conclusion

This paper presented a small-scale exploratory investigation of the Mastodon platform. Despite the increased research on SNSs with U&G Theory, most studies have focused on mainstream media. This study's main contribution to social science research is to discover the reasons behind the rise of ASMs such as Mastodon and provide a new way to think about this new SNS. The present study fills the ASM inspection gap to examine the G.S. and G.O. from the U&G perspective. Moreover, to better understand the G.S. and G.O.'s differences, the comparison approach was applied. This approach was based on many satisfaction studies, e.g., (Gibbs, O'Reilly, & Brunette, 2014; Johnson & Yang, 2009).

The survey results (Table 4) show that the mean for the 4 G.S. factors (e.g., Convenience; Privacy; Information Seeking; Social Escapism and Support) were over 3.00 on a scale of 5.00, suggesting the user's expectation of those factors for joining Mastodon is higher than "Neutral" but lower than "Agree". 3 G.S. factors (e.g., Socialization; Adult Content Sharing; Adult Content Consuming) were less than 2.50 on a scale of 5.00, suggesting the user's expectation on those factors for joining Masto-don is lower than "Neutral" but higher than "Disagree". 4 G.O. factors (e.g., Convenience; Privacy; Social Escapism and Support; Information Seeking) were over 3.00 on a scale of 5.00, suggesting the user's gratifications on those factors for using Mastodon is higher than "Neutral" but lower than "Agree". 3 G.O. factors (e.g., Adult Content Sharing; Socialization; Adult Content Consuming) were less than 2.50 on a scale of 5.00, suggesting the user's expectation on those factors for joining Mastodon is lower than "Neutral" but higher than "Disagree".

Of the 7 G.S. and G.O. pairs tested (Table 5), "Adult Content Sharing" and "Adult Content Consuming" presented statistically significant mean differences, suggesting that users gained gratification from using Mastodon more than expected. "Social Escapism and Support" presented significant mean differences, suggesting users' expectations were met. 4 pairs (e.g., Socialization; Privacy; Convenience; Information Seeking) were not statistically significant due to similar G.S. and G.O. factor scores, suggesting that users expected to gain those gratifications from using Mastodon in the beginning and expectations were met. However, consider that both factors' scores of "Socialization" were around 2.50 (e.g., G.S. = 2.44835; G.O. = 2.51423), which means user's satisfaction was lower than "Neutral" but higher than "Disagree". Therefore, the most satisfying reasons for users joining and using Mastodon were "Information Seeking", "Convenience" and "Privacy". It suggests that joining Mastodon's main expectations are the platform is informative and easy to use, and the privacy setting will shelter users' information from data mining. And similar gratifications are received from using the platform. Users are satisfied with the ongoing use of the Mastodon.

"Adult Content Sharing" and "Adult Content Consuming" were surprisingly lower than 2.50. During the online conversation with a Mastodon user, the two factors and related items were always mentioned by participants. As one of the components of social media, adult content (e.g., pornography) has been recognized as a visual stimulus (Luscombe, 2016), and the combination of computer access, sexual pleasure, and the brain's mechanisms for learning could make online pornography acutely habit-forming. Zignani et al. (Zignani, Quadri, Galdeman, Gaito, & Rossi, 2019) have addressed this in the research: Since each instance defines its policy and its members' code of conduct, it is worth noting that pornography or content for an adult audience may not be prohibited in some instances. 8.9% of posts are labeled as "sensitive". It is such a great percentage for such a small-scale ASM. Muntinga et al. (Muntinga, Moorman, & Smit, 2011) have also addressed: entertainment motivation covers several media gratifications related to escaping or being diverted from problems or routine; emotional release or relief; relaxation; cultural or aesthetic enjoyment; passing the time; and Adult Content Consuming. Hence, in this research, "Adult Content Sharing" and "Adult Content Consuming" would be listed as gratification factors and gained high scores to some extent. However, the opposite results were obtained. The plausible explanation is that respondents would lie about the answers because lying would look more attractive or for privacy or protection concerns (Drouin, Miller, Wehle, & Hernandez, 2016), even if it is an anonymous online survey.

The most impressive finding from the analysis is that the users are seeking "Convenience" as the main form of gratification more than "Privacy" (Table 4), which is the unique function of the D.W. structure. The gratification of "Convenience" is assessed using three items modified from Bae (Bae, 2018). It suggests that users value the features of "simple, handy, and easy to control" more than functional capability. Another plausible explanation is that users value the advertisement-free platform that Mastodon provides. All the instances have a clean and ad-free timeline and present a spotless, simple, and friendly interface, giving users more convenience. That explained why users are satisfied with that factor (Table 5).

Additionally, during the investigation, the researcher found that the number of active Mastodon users was considerably lower than mainstream media (e.g., Twitter), which is reasonable because ad-free means no marketers will support the platform. The technical logic behind advertising was based on user data collecting. ASM refused that. As Gehl (Gehl, 2015) mentioned, the refusal of advertising has consequences: the ASM does not give in to the technical, infrastructural, or organizational demands that marketers would make upon them.

"Privacy" is the second satisfying reason for people joining and using Mastodon. It is understandable because SNS Users always demonstrate strong privacy concerns online (Buchanan, Paine, Joinson, & Reips, 2007; Young, A. L. & Quan-Haase, 2009). As a result, the Mastodon returned to individual users or groups of users not only their data but also the control over their contents (Zignani et al., 2019). However, many researchers, e.g., (Debatin, Lovejoy, Horn, & Hughes, 2009; Tufekci, 2008), found little to no relationship between online privacy concerns and information disclosure on online social network sites. It is true because people who have privacy concerns would prefer not to use any SNS in their life. Tufekci (Tufekci, 2008) has conducted: "We tested to see whether those expressing higher degrees of privacy concerns were less likely to use social network sites. Indeed, nonusers of social network sites had higher levels of privacy concerns." As a factor for measuring gratifications, "Privacy" is assessed using five items modified from Smith et al. (Smith, Milberg, & Burke, 1996), Buchanan et al. (Buchanan et al., 2007), Heravi et al. (Heravi, Mubarak, & Choo, 2016), and Dinev et al. (Dinev & Hart, 2004). Two items summarized from the Mastodon user's online conversion: "None-decentralized SNSs cannot provide sufficient privacy settings" and "Avoid being judged by sharing adult content".

"Social Escapism and Social Support" were two separate factors modified from Bae (Bae, 2018) . These items were combined during factor analysis. On the one hand, Korgaonkar et al. (Korgaonkar & Wolin, 1999) have defined social escapism as a "pleasurable, fun and enjoyable factor". Hastall (Hastall, 2017) found escapism "is just one of several ways in which individuals deal with challenging life situations". On the other hand, the SNS users are seeking emotional support at the same time. Online support provides a convenient connection with others in similar circumstances as Hwang et al. (Hwang et al., 2010) have conducted: the ability to communicate anonymously, reciprocity of social support, and a judgment-free space for people to share information about their health status. Hence, it is reasonable to mix the two concepts.

The present research has limitations:

1. The number of participants was too small: 254 responses but only 150 valid responses in this case. According to Punch et al. (Punch & Oancea, 2014), the average number for the exploratory study should be 200 participants.

2. Both G.S. and G.O. measurements could be modified. If we can use more appropriate items via an official interview with Mastodon's users, the results would be more accurate.

3. The measurement could be expanded by adding demographic information because we were surprised 78% of participants were female. Research about gender and social media by Wang et al. (Wang, Fink, & Cai, 2008) found that females prefer to use SNS to satisfy their lack of family relationships while males join SNS for removing their feeling of loneliness. Other researchers (Cho et al., 2003; Choi, 2000; Karimi et al., 2014) applied demographic factors (e.g., time consumption on SNS, education level, age, gender) to investigate the daily usage of SNS and use this information to predict motivations for media usage.

4. Lastly, the research is depended on the survey instrument's effectiveness and the subjects' ability to answer the questions accurately. This limitation is similar to most surveys and is an inherent limitation of U&G research.

Despite the increased research on SNSs applying the U&G theory, most studies have focused on mainstream social media. The present investigation was the first to examine ASM from the U&G theory perspective through exploring gratifications sought and gratifications obtained. The present study fills the gap between ASM and U&G theory. We discover why Mastodon arose and provide a new measurement, privacy, which measures users' privacy concerns to a centralized social media. The findings of this research indicated that users would join and use a social media application if it is a privacy-guaranteed platform. A better understanding of Alternative Social Media will make mainstream social media rethink their current design decisions.